\documentclass[12pt]{article}

\usepackage{amsmath}
\usepackage{graphicx}

\newcommand{\ket}[1]{| #1 \rangle}
\newcommand{\bra}[1]{\langle #1 |}
\newcommand{\hcs}[1]{#1^\dagger #1}
\newcommand{\expv}[1]{\langle #1 \rangle}
\newcommand{\pc}{\text{PC}}
\newcommand{\qc}{\text{QC}}
\newcommand{\qpc}{\text{QPC}}
\newcommand{\qqc}{\text{QQC}}

\begin{document}

\begin{center}
{\Large\bf Information, fidelity, and reversibility
in photodetection processes}
\vskip .6 cm
Hiroaki Terashima
\vskip .4 cm
{\it Department of Physics, Faculty of Education, Gunma University, \\
Maebashi, Gunma 371-8510, Japan}
\vskip .6 cm
\end{center}

\begin{abstract}
Four types of photon counters are discussed in terms of
information, fidelity, and physical reversibility:
conventional photon counter, quantum counter,
and their quantum nondemolition (QND) versions.
It is shown that when a photon field to be measured is in
an arbitrary superposition of vacuum and one-photon states,
the quantum counter is the most reversible,
the QND version of conventional photon counter
provides the most information,
and the QND version of quantum counter causes the smallest state change.
Our results suggest that the physical reversibility of a counter
tends to decrease the amount of information obtained by the counter.
\end{abstract}

\begin{flushleft}
{\footnotesize
}
\end{flushleft}

\section{Introduction}
When a quantum measurement provides information
about a physical system, it inevitably changes
the state of the system into another state
via non-unitary state reduction.
This property is of great interest 
not only in the foundations of quantum mechanics
but also in quantum information processing
and communication~\cite{NieChu00}, e.g.,
in quantum cryptography~\cite{BenBra84,Ekert91,Bennet92,BeBrMe92}.
However, such a state change by measurement
is not necessarily irreversible~\cite{UedKit92,UeImNa96},
despite being
widely believed to be intrinsically irreversible~\cite{LanLif77}.
A quantum measurement is said to
be physically reversible~\cite{UeImNa96,Ueda97}
if the pre-measurement state can be recovered from
the post-measurement state
with a nonzero probability of success
by means of a second measurement,
referred to as reversing measurement.
Recently, physically reversible measurements have been proposed with
various systems~\cite{Imamog93,Royer94,TerUed05,KorJor06,%
TerUed07,SuAlZu09,XuZho10} and discussed
in the context of quantum computation~\cite{KoaUed99,TerUed03},
and have been experimentally demonstrated
using a superconducting phase qubit~\cite{KNABHL08}
and a photonic qubit~\cite{KCRK09}.
Therefore, it would be worth discussing
the state change by a measurement
together with its physical reversibility.

The necessary and sufficient condition for physical reversibility
is that the operator $\hat{M}$ describing the state change by the measurement
has a bounded left inverse $\hat{M}^{-1}$~\cite{UeImNa96,Ueda97}.
In fact, to recover the pre-measurement state,
the reversing measurement is constructed so that
it applies $\hat{M}^{-1}$ to the measured system
to cancel the effect of $\hat{M}$
when a preferred outcome is obtained.
Interestingly, the reversing measurement completely
erases the information provided by the first measurement
when it successfully recovers the pre-measurement state
(see Erratum of Ref.~\cite{Royer94}),
although a physically reversible measurement actually provides
some information about the measured system
in contrast to
the unitarily reversible measurements~\cite{MabZol96,NieCav97}.
Therefore, a reversing operation based on $\hat{M}^\dagger$,
instead of $\hat{M}^{-1}$,
has been proposed~\cite{TerUed07b},
which can approximately recover the pre-measurement state
especially with increasing, rather than decreasing, information gain
for a weak measurement.
Further discussions
of information gain by physically reversible measurement
can be seen in other studies~\cite{Ban01,DArian03}.

In this article, we investigate four types of photon counters
to compare them in terms of information gain, state change,
and physical reversibility of the photodetection processes.
The first counter is a conventional photon counter
that operates by absorption of photons, and
the second counter is a quantum counter~\cite{Bloemb59,Mandel66}
that operates by stimulated emission of photons.
The third and fourth counters are
the quantum nondemolition (QND)~\cite{BraKha96} versions
of the first and second counters, that is,
the QND photon and QND quantum counters,
which perform unsharp measurements of photon number
without perturbing photon-number states.
Among the four counters,
quantum counter and its QND version are physically reversible.
For each counter,
we evaluate the amount of information gain
using a decrease in Shannon entropy~\cite{DArian03,TerUed07b},
the degree of state change using fidelity~\cite{Uhlman76},
and the degree of physical reversibility
using the maximal successful probability
of reversing measurement~\cite{KoaUed99},
assuming that a photon field to be measured is in
an arbitrary superposition of vacuum and one-photon states.

This article is organized as follows:
Section~\ref{sec:QM} reviews
a mathematical formulation of quantum measurement and
the physical reversibility in quantum measurement.
Sections~\ref{sec:PC}, \ref{sec:QC}, \ref{sec:QPC}, and \ref{sec:QQC}
discuss the conventional photon counter, quantum counter, QND photon counter,
and QND quantum counter, respectively,
calculating the information gain, fidelity,
and physical reversibility in a two-state model.
Section~\ref{sec:summary} summarizes our results,
compares the four counters,
and discusses an implementation of a QND quantum counter
proposed in this article.

\section{\label{sec:QM}Quantum Measurement}
Here, we briefly review a mathematical formulation of
quantum measurement together with its physical reversibility.
Let $\ket{\psi}$ be an unknown pre-measurement state
of a system to be measured.
To obtain information about the state,
we perform an indirect measurement using a probe as follows.
We first prepare the probe in a state $\ket{i}_\text{p}$
and then turn on an interaction between the probe and the system
via an interaction Hamiltonian $\hat{H}_\text{int}$
during a time interval $\Delta t$.
After the interaction,
the state of the whole system becomes
$\hat{U}_\text{int}\ket{\psi}\ket{i}_\text{p}$,
where $\hat{U}_\text{int}=\exp\left(-i\hat{H}_\text{int}\Delta t/\hbar\right)$.
Finally,
we perform a projective measurement on the probe with respect to
an orthonormal basis $\{\ket{m}_\text{p}\}$.
From the outcome $m$, we can indirectly obtain
some information about the state.
Below we shall show what and how much information we can obtain
in the case of photodetection processes.

The measurement yields an outcome $m$
with probability
\begin{equation}
 p_m=\bra{\psi} \hat{M}_m^\dagger\hat{M}_m \ket{\psi},
\label{eq:pm}
\end{equation}
where $\hat{M}_m={}_\text{p}\bra{m}\hat{U}_\text{int}\ket{i}_\text{p}$,
and simultaneously changes the state of the system from $\ket{\psi}$ into
\begin{equation}
\ket{\psi_m}=\frac{1}{\sqrt{p_m}} \hat{M}_m \,\ket{\psi},
\label{eq:psim}
\end{equation}
depending on the outcome $m$.
In other words,
a quantum measurement is mathematically
described by a set of
linear operators $\{\hat{M}_m\}$~\cite{DavLew70,NieChu00},
called measurement operators, that satisfy the completeness condition
\begin{equation}
\sum_m\hcs{\hat{M}_m}=\hat{I},
\label{eq:compcond}
\end{equation}
where $\hat{I}$ is the identity operator.
The probability and post-measurement state are then given for each outcome $m$
by Eqs.~(\ref{eq:pm}) and (\ref{eq:psim}), respectively.
Conversely,
for a give set of linear operators $\{\hat{M}_m\}$ satisfying
the completeness condition (\ref{eq:compcond}),
an indirect measurement described by $\{\hat{M}_m\}$ can always
be constructed by choosing the initial state $\ket{i}_\text{p}$,
the interaction $\hat{U}_\text{int}$,
and the orthonormal basis $\{\ket{m}_\text{p}\}$ of the probe.

Although the measurement changes the state of the system
as in Eq.~(\ref{eq:psim}),
this state change is physically reversible if and only if
$\hat{M}_m$ has a bounded left inverse~\cite{UeImNa96,Ueda97}.
In fact, to undo the state change,
consider performing another measurement, called reversing measurement, on
the post-measurement state (\ref{eq:psim}).
The reversing measurement is described by a set of
measurement operators $\{\hat{R}^{(m)}_\nu\}$
that satisfy
\begin{equation}
\sum_\nu\hat{R}^{(m)\dagger}_\nu\hat{R}^{(m)}_\nu=\hat{I}
\label{eq:revunit}
\end{equation}
and for a particular $\nu_0$,
\begin{equation}
 \hat{R}^{(m)}_{\nu_0}=\eta_m\, \hat{M}_m^{-1}
\end{equation}
with a complex constant $\eta_m$.
The index $\nu$ denotes the outcome of the reversing measurement.
Therefore, if the reversing measurement yields the particular outcome $\nu_0$,
it restores the pre-measurement state $\ket{\psi}$
except for an overall phase factor
from Eq.~(\ref{eq:psim}) as
\begin{equation}
  \ket{\psi_{m\nu_0}}=\frac{1}{\sqrt{p_{m\nu_0}}}\,
           \hat{R}^{(m)}_{\nu_0} \,\ket{\psi_m} \propto \ket{\psi},
\end{equation}
where
\begin{equation}
 p_{m\nu_0}=\bra{\psi_m} \hat{R}^{(m)\dagger}_{\nu_0}\hat{R}^{(m)}_{\nu_0}
   \ket{\psi_m}=\frac{|\eta_m|^2}{p_m}
\label{eq:condrev}
\end{equation}
is the probability for the second outcome $\nu_0$ given the first outcome $m$,
and thus is the successful probability of the reversing measurement.
Since the completeness condition (\ref{eq:revunit}) requires
$\bra{\psi'}\hat{R}^{(m)\dagger}_{\nu_0}\hat{R}^{(m)}_{\nu_0}\ket{\psi'}\le 1$
for any state $\ket{\psi'}$,
the upper bound for $|\eta_m|^2$ becomes~\cite{KoaUed99}
\begin{equation}
|\eta_m|^2\le
\inf_{\ket{\psi'}}\,\bra{\psi'} \hat{M}_m^\dagger\hat{M}_m \ket{\psi'}
\equiv b_m,
\label{eq:upper}
\end{equation}
which does not depend on the pre-measurement state $\ket{\psi}$.
The upper bound $b_m$ is called the background of $\hat{M}_m$,
implying that the measurement $\{\hat{M}_m\}$
yields the outcome $m$ with a probability not less than $b_m$ for any state.
Combining Eqs.~(\ref{eq:condrev}) and (\ref{eq:upper}), we find that
if the pre-measurement state is $\ket{\psi}$ and
the first outcome is $m$,
the maximal successful probability of the reversing measurement is given by
\begin{equation}
  R\bigl(m,\ket{\psi}\bigr)\equiv \max_{\eta_m}\, p_{m\nu_0}
    =\frac{b_m}{p_m}.
\label{eq:maxrecover}
\end{equation}
That is, we can, \emph{in principle},
recover the unknown pre-measurement state $\ket{\psi}$
from the post-measurement state $\ket{\psi_m}$
with the probability (\ref{eq:maxrecover}),
even though it would be difficult to experimentally implement
the reversing measurement $\{\hat{R}^{(m)}_\nu\}$
with $|\eta_m|^2=b_m$ as an indirect measurement.

\section{\label{sec:PC}Photon Counter}
A photon counter usually detects photons one by one
from a photon field.
This means that the photon counter detects at most one photon
during a short time interval.
When detecting one photon (``one-count'' process),
the counter annihilates the detected photon
from the photon field.
Even in the case when no photon is detected (``no-count'' process),
the counter changes the state of the photon field
owing to the obtained information that no photon was detected
during the time interval.
A physical model of the photon counter is described
in accordance with the indirect measurement in Sec.~\ref{sec:QM}.
In this case, the probe is a two-level atom having
a ground state $\ket{g}_\text{p}$
and an excited state $\ket{e}_\text{p}$
with a raising operator $\hat{\sigma}_+=\ket{e}_{\text{p}\,\text{p}}\bra{g}$
and a lowering operator $\hat{\sigma}_-=\ket{g}_{\text{p}\,\text{p}}\bra{e}$.
The initial state of the atom is the ground state $\ket{g}_\text{p}$,
and the interaction Hamiltonian between
the atom and the photon field is the Jaynes-Cummings Hamiltonian
\begin{equation}
\hat{H}_\text{int}=\hbar g \left(\hat{a} \hat{\sigma}_+
    + \hat{a}^\dagger\hat{\sigma}_- \right),
\label{eq:Jaynes-Cummings}
\end{equation}
where $g$ is a coupling constant, and
$\hat{a}^\dagger$ and $\hat{a}$ are 
the creation and annihilation operators of the photon.
The projective measurement on the atom is
with respect to
the basis $\{\ket{g}_\text{p},\ket{e}_\text{p}\}$.
As a result of the measurement,
if the atom is found to be in the excited state $\ket{e}_\text{p}$,
we recognize that the one-count process has occurred
with the absorption of a photon.
On the other hand,
if the atom is found to be still in the ground state $\ket{g}_\text{p}$,
we recognize that the no-count process has occurred
with detecting no photon.

In terms of the measurement operator in Sec.~\ref{sec:QM},
the one- and no-count processes are described
by~\cite{SriDav81,UeImOg90,UeImNa96},
\begin{equation}
 \hat{M}_1=\gamma \hat{a}, \qquad
 \hat{M}_0\simeq\hat{I}-\frac{\gamma^2}{2} \hat{a}^\dagger \hat{a},
\label{eq:PC-operator}
\end{equation}
respectively, where $\gamma= g\Delta t$ is a constant
that is assumed to be so small that
we can ignore the fourth and higher order terms in $\gamma$.
In fact, the annihilation operator in $\hat{M}_1$
annihilates a photon from the photon field through
the state reduction (\ref{eq:psim}) in the one-count process.
Moreover, combined with $\hat{M}_1$,
the measurement operator $\hat{M}_0$ for the no-count process
satisfies the completeness condition (\ref{eq:compcond}), i.e.,
\begin{equation}
\sum_{m=0, 1} \hcs{\hat{M}_m}\simeq \hat{I}
\label{eq:PC-complete}
\end{equation}
up to the order of $\gamma^3$.
This means that we can regard the one-count and no-count processes
as a mutually exclusive and exhaustive set of events
in the measurement.

\subsection{General Model}
To evaluate the amount of information provided by the photon counter,
we assume that the pre-measurement state of the photon field is known to
be one of the predefined pure states $\{\ket{\psi(a)}\}$
with equal probability, $p(a)=1/N$, where $a=1,\ldots,N$,
although the pre-measurement state is unknown.
Because in quantum measurement the pre-measurement state is usually
an arbitrary unknown state,
the set $\{\ket{\psi(a)}\}$ is essentially an infinite set ($N\to\infty$)
to cover the Hilbert space of the photon field.
Each state can be expanded by
the eigenstates $\{\ket{n}\}$ of 
the photon-number operator $\hat{a}^\dagger\hat{a}$ as
\begin{equation}
  \ket{\psi(a)}=\sum_n c_n(a)\, \ket{n}
\end{equation}
with $n=0,1,2,\ldots$, and
the coefficients $\{c_n(a)\}$ that obey the
normalization condition $\sum_n \left| c_n(a)\right|^2=1$.
Our lack of information about the photon field can be quantified by
the Shannon entropy associated
with the probability distribution $\{p(a)\}$ as
\begin{equation}
  H_0=-\sum_a p(a)\log_2 p(a)=\log_2 N.
\end{equation}
Next, we perform a measurement
by the photon counter (\ref{eq:PC-operator})
to obtain a piece of information about the photon field.
According to Eq.~(\ref{eq:pm}),
if the pre-measurement state is $\ket{\psi(a)}$,
the one-count process occurs with probability
\begin{equation}
 p^\pc (1|a) = \bra{\psi(a)} \hat{M}_1^\dagger\hat{M}_1 \ket{\psi(a)}
    = \gamma^2 n_1(a),
\label{eq:PC-probability}
\end{equation}
where
\begin{equation}
   n_1(a) \equiv \sum_n n \left| c_n(a)\right|^2.
\end{equation}
Since the probability for $\ket{\psi(a)}$ is $p(a)=1/N$,
the total probability for the one-count process is given by
\begin{equation}
 p^\pc(1) =\sum_a  p^\pc(1|a)\,p(a)
      =\frac{1}{N}\sum_a \gamma^2 n_1(a)
      =\gamma^2 \overline{n_1},
\label{eq:PC-total}
\end{equation}
where the overline denotes the average over $a$,
\begin{equation}
   \overline{f} \equiv \frac{1}{N}\sum_a f(a).
\end{equation}
On the contrary, given that the photon counter detects one photon,
we can find the probability for
the pre-measurement state $\ket{\psi(a)}$ as
\begin{equation}
 p^\pc(a|1)=\frac{p^\pc(1|a)\,p(a)}{p^\pc(1)}
           =\frac{n_1(a)}{N\overline{n_1}}
\label{eq:PC-conditional}
\end{equation}
from Bayes' rule.
Using this probability distribution,
our lack of information after the one-count process
is evaluated by the Shannon entropy as follows:
\begin{equation}
  H^\pc(1) =-\sum_a p^\pc(a|1)\log_2 p^\pc(a|1)=\log_2 N
    -\frac{\overline{n_1\log_2 n_1}-
           \overline{n_1}\log_2\overline{n_1}}{\overline{n_1}}.
\end{equation}
The information gain by the one-count process
is then defined by the decrease in Shannon entropy as
\begin{equation}
I^\pc(1)=H_0-H^\pc(1)=
 \frac{\overline{n_1\log_2 n_1}-
           \overline{n_1}\log_2\overline{n_1}}{\overline{n_1}},
\label{eq:PC-information}
\end{equation}
which does not depend on $\gamma$
(i.e., on the coupling constant $g$
between the photon counter and the photon field).
That is,
this information gain is a measure of
how much our knowledge about the pre-measurement state increases
when we revise the probability distribution
from $p(a)=1/N$ to $p^\pc(a|1)$ according to the outcome.
Note that it results from
a \emph{single} measurement outcome~\cite{DArian03,TerUed07b}
without averaging all the outcomes,
and that it indicates the state of the pre-measurement
rather than a value of some observable.
Similar to the one-count process,
we obtain the total probability for the no-count process which
is given as $p^\pc (0) \simeq 1-\gamma^2 \overline{n_1}$;
this information gain by the no-count process is $I^\pc(0)\simeq 0$
up to the order of $\gamma^3$.
Therefore, averaging over the outcomes $m=0,1$,
we find that the mean information gain by the measurement is given by
\begin{equation}
  I^\pc=\sum_m p^\pc(m) \,I^\pc(m)
       \simeq \gamma^2\left(\,
     \overline{n_1\log_2 n_1}-
           \overline{n_1}\log_2\overline{n_1}\,\right),
\label{eq:PC-I}
\end{equation}
which is identical to
the mutual information~\cite{NieChu00}
of the random variables $\{a\}$ and $\{m\}$:
\begin{equation}
   I^\pc=\sum_{m,a} p^\pc(a|m)\,p^\pc(m)\, \log_2\frac{p^\pc(a|m)}{p(a)}.
\end{equation}

Unfortunately,
the measurement changes the state of the photon field.
The state change can be evaluated by the fidelity~\cite{Uhlman76,NieChu00}
between the pre-measurement and post-measurement states.
According to Eq.~(\ref{eq:psim}),
when the pre-measurement state is $\ket{\psi(a)}$,
the post-measurement state after the one-count process is
\begin{equation}
   \ket{\psi(1,a)}^\pc 
    = \frac{1}{\sqrt{p^\pc(1|a)}}\,\hat{M}_1 \ket{\psi(a)}
    =\frac{1}{\sqrt{n_1(a)}}\sum_n\sqrt{n+1}\, c_{n+1}(a) \, \ket{n},
\end{equation}
whose fidelity to $\ket{\psi(a)}$ is
\begin{equation}
  F^\pc(1,a) = \bigl|\expv{\psi(a)|\psi(1,a)}^\pc\bigr|
     =\frac{1}{\sqrt{n_1(a)}}
     \left|\sum_n \sqrt{n+1}\,c_n^\ast(a)\,c_{n+1}(a)\right|.
\end{equation}
Since the index $a$ is unknown,
we average over $a$ with the probability (\ref{eq:PC-conditional})
to obtain the fidelity after the one-count process as
\begin{equation}
 F^\pc(1) =\sum_a p^\pc(a|1)\, F^\pc(1,a)
     = \overline{\frac{\sqrt{n_1}}{\overline{n_1}}
     \left|\sum_n \sqrt{n+1}\,c_n^\ast\,c_{n+1}\right|}.
\label{eq:PC-fidelity}
\end{equation}
On the other hand,
the fidelity after the no-count process
is $F^\pc(0)\simeq 1$ up to the order of $\gamma^3$.
The mean fidelity after the measurement is thus given by
\begin{equation}
  F^\pc=\sum_m p^\pc(m)\, F^\pc(m)\simeq 1-\gamma^2 \overline{n_1}
     +\gamma^2 \overline{\sqrt{n_1}
     \left|\sum_n \sqrt{n+1}\,c_n^\ast\,c_{n+1}\right|}.
\label{eq:PC-F}
\end{equation}

We can, however, undo this state change of the photon field
if the measurement is physically reversible
as described in Sec.~\ref{sec:QM}.
The physical reversibility
can be evaluated by the maximal successful probability (\ref{eq:maxrecover})
of the reversing measurement.
If the pre-measurement state is $\ket{\psi(a)}$ and
the outcome is the one-count process,
it becomes
\begin{equation}
  R^\pc(1,a) = \frac{b^\pc(1)}{p^\pc(1|a)}
             = \frac{n_{1\text{i}}}{n_1(a)},
\label{eq:PC-reverpsi}
\end{equation}
where $b^\pc(1)$ is the background of $\hat{M}_1$
defined in Eq.~(\ref{eq:upper}), namely,
\begin{equation}
b^\pc(1)= \inf_{a'} p^\pc(1|a')
=\gamma^2\inf_{a'} n_1(a')\equiv \gamma^2 n_{1\text{i}}.
\end{equation}
Averaging over $a$ with the probability (\ref{eq:PC-conditional}),
we find the reversibility of the one-count process as
\begin{equation}
    R^\pc(1) = \sum_a p^\pc(a|1)\,R^\pc(1,a)
             = \frac{n_{1\text{i}}}{\,\overline{n_1}\,}.
\label{eq:PC-reversibility}
\end{equation}
Similarly, using the background of $\hat{M}_0$,
\begin{equation}
b^\pc(0)= \inf_{a'} p^\pc(0|a')\simeq
1-\gamma^2 \sup_{a'} n_1(a')\equiv
1-\gamma^2 n_{1\text{s}},
\end{equation}
the reversibility of the no-count process is found to be
\begin{equation}
  R^\pc(0,a) = \frac{b^\pc(0)}{p^\pc(0|a)}
      \simeq \frac{1-\gamma^2 n_{1\text{s}}}{1-\gamma^2n_1(a)}
\end{equation}
if the pre-measurement state is $\ket{\psi(a)}$, and is
\begin{equation}
 R^\pc(0) = \sum_a p^\pc(a|0)\,R^\pc(0,a)
   \simeq \frac{1-\gamma^2 n_{1\text{s}}}{1-\gamma^2\overline{n_1}}
\label{eq:PC-reversibility0}
\end{equation}
if averaged over $a$.
The mean reversibility of the measurement thus becomes
\begin{equation}
  R^\pc=\sum_m p^\pc(m)\, R^\pc(m)
   \simeq
   1-\gamma^2\left(n_{1\text{s}}-n_{1\text{i}} \right).
\label{eq:PC-R}
\end{equation}
It is easy to check from Eqs.~(\ref{eq:PC-conditional}),
(\ref{eq:PC-reverpsi}), and (\ref{eq:PC-reversibility}) that
\begin{equation}
    R^\pc=\sum_m \inf_{a'} p^\pc(m|a').
\end{equation}
That is, the quantity (\ref{eq:PC-R}) is identical to
the degree of physical reversibility of measurement
discussed by Koashi and Ueda~\cite{KoaUed99}.

\subsection{Two-state Model}
As an example, we consider a situation where
the photon field is in an arbitrary superposition
of the states $\ket{0}$ and $\ket{1}$.
That is, the set of predefined states $\{\ket{\psi(a)}\}$
consists of all possible states of the form
\begin{equation}
  \ket{\psi(a)}=\cos\frac{\theta}{2}\,\ket{0}+
    e^{i\phi}\sin\frac{\theta}{2}\,\ket{1},
\label{eq:psi}
\end{equation}
where $0\le\theta\le\pi$ and $0\le\phi<2\pi$.
The index $a$ now represents
the two continuous angles $(\theta,\phi)$.
Therefore, the probability $p(a)=1/N$ is replaced with
a probability \emph{density} $p(a)=1/4\pi$
using the volume element $\sin \theta d\theta d\phi$
and the summation over $a$ is replaced
with an integral over $(\theta,\phi)$, namely,
\begin{equation}
   \frac{1}{N}\sum_a   \quad\longrightarrow\quad
     \frac{1}{4\pi}\int^{2\pi}_0 d\phi\,
     \int^\pi_0 d\theta \sin \theta.
\end{equation}

\begin{figure}
\begin{center}
\includegraphics[scale=0.7]{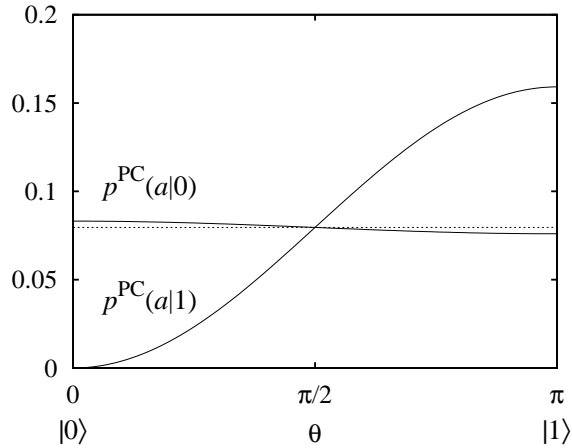}
\end{center}
\caption{\label{fig1}
Probability density for the pre-measurement state $\ket{\psi(a)}$
conditioned by the one-count process $p^\pc(a|1)$
and that conditioned by the no-count process $p^\pc(a|0)$
as functions of $\theta$ when $\gamma=0.3$.
The dotted line indicates the initial probability density $p(a)=1/4\pi$.}
\end{figure}

If the pre-measurement state is $\ket{\psi(a)}$,
the probability (\ref{eq:PC-probability}) for the one-count process is
\begin{equation}
   p^\pc (1|a) =\gamma^2 \sin^2\frac{\theta}{2},
\end{equation}
since for the state $\ket{\psi(a)}$ in Eq.~(\ref{eq:psi}) we have
\begin{equation}
   n_1(a)=\sum_{n=0,1} n \left| c_n(a)\right|^2
     =\left| c_1(a)\right|^2=\sin^2\frac{\theta}{2}.
\label{eq:n1a}
\end{equation}
The total probability (\ref{eq:PC-total}) for the one-count process
then becomes
\begin{equation}
   p^\pc (1) =\frac{1}{2}\gamma^2,
\end{equation}
because of
\begin{equation}
  \overline{n_1}=\frac{1}{N}\sum_a n_1(a)
    =   \frac{1}{4\pi}\int^{2\pi}_0 d\phi\,
     \int^\pi_0 d\theta \sin \theta\times  \sin^2\frac{\theta}{2}
    =\frac{1}{2}.
\label{eq:n1}
\end{equation}
On the contrary, given the one-count process,
the probability density (\ref{eq:PC-conditional}) for
the pre-measurement state $\ket{\psi(a)}$ is
\begin{equation}
 p^\pc(a|1)=\frac{1}{2\pi}
       \sin^2\frac{\theta}{2},
\end{equation}
while the corresponding probability density for the no-count process is
\begin{equation}
 p^\pc(a|0) \simeq \frac{1}{4\pi}\left[1-\gamma^2
       \left(\sin^2\frac{\theta}{2}-\frac{1}{2}\right)\right].
\end{equation}
These probability densities are the content of information
provided by the photon counter (\ref{eq:PC-operator}).
Figure~\ref{fig1} shows these densities as
functions of $\theta$ when $\gamma=0.3$.
Although all the states were equally probable before the measurement,
as shown by the dotted line,
the one-count process increases the possibility of $\ket{1}$
and completely excludes the possibility of $\ket{0}$,
as shown by the line $p^\pc(a|1)$.
On the contrary, the no-count process 
decreases the possibility of $\ket{1}$ and
increases the possibility of $\ket{0}$,
as shown by the line $p^\pc(a|0)$,
but so slightly that $I^\pc(0)\simeq0$.
Calculating
\begin{align}
 \overline{n_1\log_2 n_1}
    &=\frac{1}{N}\sum_a n_1(a)\log_2 n_1(a) \notag \\
    &=\frac{1}{4\pi}\int^{2\pi}_0 d\phi\,
     \int^\pi_0 d\theta \sin \theta\times  \sin^2\frac{\theta}{2}
        \log_2 \sin^2\frac{\theta}{2} =-\frac{1}{4\ln 2}
  \label{eq:n1logn1}
\end{align}
with $\log_2 x=\ln x/\ln2$,
we obtain the information gain (\ref{eq:PC-information})
by the one-count process as
\begin{equation}
  I^\pc(1) =1-\frac{1}{2\ln 2}
       \simeq 0.279
\label{eq:PC-info1}
\end{equation}
and the mean information gain (\ref{eq:PC-I}) by the measurement as
\begin{equation}
  I^\pc\simeq \left(\frac{1}{2}-\frac{1}{4\ln2} \right)\gamma^2
   \simeq 0.139\gamma^2.
\end{equation}
Furthermore,
the fidelity (\ref{eq:PC-fidelity}) after the one-count process becomes
\begin{align}
   F^\pc(1) &= \overline{\frac{\sqrt{n_1}}{\overline{n_1}}
     \bigl|\,c_0^\ast\,c_1\bigr|}
     = \frac{1}{N}\sum_a \frac{\sqrt{n_1(a)}}{\overline{n_1}}
     \bigl|\,c_0^\ast(a)\,c_1(a)\bigr| \notag \\
    &= \frac{1}{4\pi}\int^{2\pi}_0 d\phi\,
     \int^\pi_0 d\theta \sin \theta\times 2\sin^2\frac{\theta}{2}
        \cos\frac{\theta}{2}=\frac{8}{15}
\label{eq:PC-fide1}
\end{align}
and the mean fidelity (\ref{eq:PC-F}) after the measurement becomes
\begin{equation}
  F^\pc  \simeq 1-\frac{7}{30}\gamma^2.
\end{equation}
Since
$n_{1\text{i}}=\inf_{a'} n_1(a')=0$ with $\ket{\psi(a')}=\ket{0}$
and $n_{1\text{s}}=\sup_{a'} n_1(a')=1$ with $\ket{\psi(a')}=\ket{1}$,
the reversibilities (\ref{eq:PC-reversibility})
and (\ref{eq:PC-reversibility0}) of the one-count and no-count processes
are given by
\begin{align}
  R^\pc(1) &=0,  \label{eq:PC-reversibility1} \\
  R^\pc(0) &\simeq 1-\frac{1}{2}\gamma^2,
\end{align}
respectively.
The mean reversibility (\ref{eq:PC-R}) of the measurement is thus
\begin{equation}
  R^\pc\simeq 1-\gamma^2.
\end{equation}
From Eq.~(\ref{eq:PC-reversibility1}), we can see that
the one-count process of the photon counter (\ref{eq:PC-operator})
is not physically reversible.
This means that we can never recover the pre-measurement state
from the post-measurement state unless we know the pre-measurement state.
The irreversibility originates from the fact that
the photon counter does not respond to the vacuum state~\cite{UedKit92},
namely, $p^\pc(1|a)=0$ for $\ket{\psi(a)}=\ket{0}$.

\section{\label{sec:QC}Quantum Counter}
A quantum counter is a photon counter that
operates by stimulated emission,
rather than by absorption, of photons.
It was proposed to detect infrared photons~\cite{Bloemb59}
or to measure 
antinormally ordered correlation functions~\cite{Mandel66,UNSTN04},
and was discussed
to show reversibility in quantum measurement~\cite{UedKit92}.
A physical model of the quantum counter is
the indirect measurement with the two-level atom and
the Jaynes-Cummings Hamiltonian (\ref{eq:Jaynes-Cummings})
as in the photon counter in Sec.~\ref{sec:PC}.
However, in this case,
the atom is first prepared in the excited state $\ket{e}_\text{p}$.
After the interaction and
the projective measurement,
if the atom is found to be in the ground state $\ket{g}_\text{p}$,
we recognize that the one-count process has occurred
with the emission of a photon.
On the other hand,
if the atom is found to be still in the excited state $\ket{e}_\text{p}$,
we recognize that the no-count process has occurred
with detecting no photon.

The action of the quantum counter is described
by the measurement operators
for one-count and no-count processes~\cite{UedKit92,UeImNa96}
\begin{equation}
 \hat{L}_1=\gamma \hat{a}^\dagger, \qquad
 \hat{L}_0\simeq\hat{I}-\frac{\gamma^2}{2} \hat{a}\hat{a}^\dagger.
\label{eq:QC-operator}
\end{equation}
As seen from $\hat{L}_1$,
the quantum counter creates a new photon in the photon field
by stimulated or spontaneous emission
in the one-count process through the state reduction (\ref{eq:psim})
as opposed to the conventional photon counter (\ref{eq:PC-operator}).
Similar to $\hat{M}_1$ and $\hat{M}_0$,
the measurement operators $\hat{L}_1$ and $\hat{L}_0$
also satisfy the completeness condition (\ref{eq:compcond}) as
\begin{equation}
\sum_{m=0, 1} \hcs{\hat{L}_m}\simeq \hat{I},
\end{equation}
up to the order of $\gamma^3$.

\subsection{General Model}
The amount of information provided
by the quantum counter (\ref{eq:QC-operator})
can be evaluated using the predefined states $\{\ket{\psi(a)}\}$
as in Sec.~\ref{sec:PC}.
If the pre-measurement state is $\ket{\psi(a)}$,
the probability for the one-count process is
\begin{equation}
   p^\qc(1|a) =\bra{\psi(a)} \hat{L}_1^\dagger\hat{L}_1 \ket{\psi(a)}
     = \gamma^2 \left[ n_1(a)+1\right]
\label{eq:QC-probability}
\end{equation}
from Eq.~(\ref{eq:pm}).
Note that the one-count process occurs even when
the photon field is in the vacuum state, $\ket{\psi(a)}=\ket{0}$,
owing to spontaneous emission,
unlike the conventional photon counter [see Eq.~(\ref{eq:PC-probability})].
In this sense, the quantum counter is sensitive not only to photons
but also to the vacuum state.
The total probability for the one-count process is then
\begin{equation}
  p^\qc(1) =\sum_a  p^\qc(1|a)\, p(a)
        =\gamma^2 \left( \overline{n_1}+1\right).
\label{eq:QC-total}
\end{equation}
On the contrary, given the one-count process,
the probability for the pre-measurement state $\ket{\psi(a)}$ is
\begin{equation}
 p^\qc(a|1) =\frac{p^\qc(1|a)\, p(a)}{p^\qc(1)}
   =\frac{n_1(a)+1}{N\left( \overline{n_1}+1\right)}.  
\label{eq:QC-conditional}
\end{equation}
Calculating the Shannon entropy $H^\qc(1)$ associated with
this probability distribution,
we find the information gain by the one-count process as
\begin{equation}
  I^\qc(1) =H_0-H^\qc(1)=\frac{\overline{(n_1+1)\log_2 (n_1+1)}-
     (\overline{n_1}+1)\log_2(\overline{n_1}+1)}{\overline{n_1}+1}.
\label{eq:QC-information}
\end{equation}
This quantifies the increase in
our knowledge about the pre-measurement state
when we revise the probability distribution from $p(a)=1/N$ to $p^\qc(a|1)$
according to the outcome.
Similarly, the total probability for the no-count process is
$p^\qc(0)\simeq 1-\gamma^2 \left( \overline{n_1}+1\right)$ and
the information gain by the no-count process
is $I^\qc(0)\simeq 0$ up to the order of $\gamma^3$.
The mean information gain by the measurement then becomes
\begin{align}
  I^\qc &=\sum_m p^\qc(m)\,  I^\qc(m)  \notag \\ 
       &\simeq \gamma^2\left[\,
     \overline{(n_1+1)\log_2 (n_1+1)}-
           (\overline{n_1}+1)\log_2(\overline{n_1}+1)\,\right].
\label{eq:QC-I}
\end{align}

On the other hand, the state change owing to the measurement
can be evaluated by fidelity.
When the pre-measurement state is $\ket{\psi(a)}$,
the post-measurement state after the one-count process is,
from Eq.~(\ref{eq:psim}),
\begin{equation}
   \ket{\psi(1,a)}^\qc = \frac{1}{\sqrt{p^\qc(1|a)}}\, \hat{L}_1 \ket{\psi(a)}
    =\frac{1}{\sqrt{n_1(a)+1}}\sum_n\sqrt{n}\, c_{n-1}(a) \, \ket{n},
\end{equation}
whose fidelity to $\ket{\psi(a)}$ is
\begin{equation}
   F^\qc(1,a) = \bigl|\expv{\psi(a)|\psi(1,a)}^\qc\bigr|
     =\frac{1}{\sqrt{n_1(a)+1}}
     \left|\sum_n \sqrt{n}\,c_n^\ast(a)\,c_{n-1}(a)\right|.
\end{equation}
Averaging over $a$ with the probability (\ref{eq:QC-conditional}),
we find that the fidelity after the one-count process is
\begin{equation}
   F^\qc(1) =\sum_a p^\qc(a|1)\,  F^\qc(1,a)
     = \overline{\frac{\sqrt{n_1+1}}{\overline{n_1}+1}
     \left|\sum_n \sqrt{n}\,c_n^\ast\,c_{n-1}\right|}.
\label{eq:QC-fidelity}
\end{equation}
Since the fidelity after the no-count process is
$F^\qc(0) \simeq 1$ up to the order of $\gamma^3$,
the mean fidelity after the measurement is given by
\begin{equation}
  F^\qc =\sum_m p^\qc(m)\,  F^\qc(m)
   \simeq 1-\gamma^2(\overline{n_1}+1)
     +\gamma^2 \overline{\sqrt{n_1+1}
     \left|\sum_n \sqrt{n}\,c_n^\ast\,c_{n-1}\right|}.
\label{eq:QC-F}
\end{equation}

Moreover,
the reversibility of the measurement can be
evaluated by the maximal successful probability (\ref{eq:maxrecover})
of its reversing measurement.
If the pre-measurement state is $\ket{\psi(a)}$,
the reversibilities of the one-count and no-count processes are
\begin{align}
  R^\qc(1,a) &= \frac{b^\qc(1)}{p^\qc(1|a)}
                      = \frac{n_{1\text{i}}+1}{n_1(a)+1},   \\
  R^\qc(0,a) &= \frac{b^\qc(0)}{p^\qc(0|a)}
       \simeq \frac{1-\gamma^2 \left( n_{1\text{s}}+1\right)}
           {1-\gamma^2 \left[ n_1(a)+1\right]},
\end{align}
respectively, from the background $b^\qc(m) = \inf_{a'} p^\qc(m|a')$
in Eq.~(\ref{eq:upper}).
Averaging over $a$ with the probability (\ref{eq:QC-conditional}),
we obtain
\begin{align}
  R^\qc(1) &= \sum_a p^\qc(a|1)\, R^\qc(1,a)
       = \frac{n_{1\text{i}}+1}{\overline{n_1}+1} ,
   \label{eq:QC-reversibility}   \\
  R^\qc(0) &= \sum_a p^\qc(a|0)\, R^\qc(0,a)
       \simeq \frac{1-\gamma^2 \left( n_{1\text{s}}+1\right)}
           {1-\gamma^2 \left( \overline{n_1}+1\right)}.
  \label{eq:QC-reversibility0}
\end{align}
The mean reversibility of the measurement is thus given by
\begin{equation}
  R^\qc =\sum_m p^\qc(m)\,  R^\qc(m)
   \simeq 1-\gamma^2\left(n_{1\text{s}}-n_{1\text{i}} \right).
\label{eq:QC-R}
\end{equation}

\subsection{Two-state Model}
\begin{figure}
\begin{center}
\includegraphics[scale=0.7]{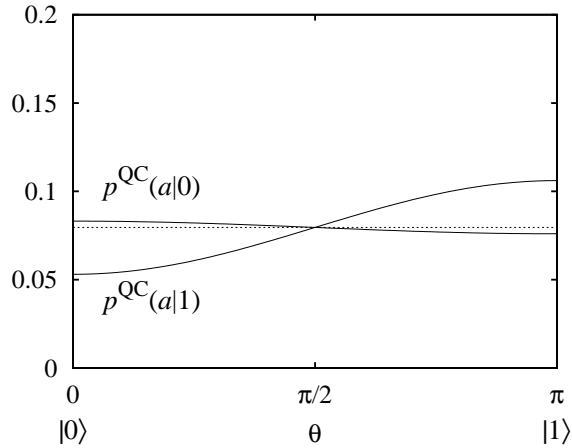}
\end{center}
\caption{\label{fig2}
Probability density for the pre-measurement state $\ket{\psi(a)}$
conditioned by the one-count process $p^\qc(a|1)$
and that conditioned by the no-count process $p^\qc(a|0)$
as functions of $\theta$ when $\gamma=0.3$.
The dotted line indicates the initial probability density $p(a)=1/4\pi$.}
\end{figure}

As an example, we consider the situation discussed
in Sec.~\ref{sec:PC}.
If the pre-measurement state is $\ket{\psi(a)}$,
the probability (\ref{eq:QC-probability}) 
for the one-count process is
\begin{equation}
   p^\qc(1|a) = \gamma^2 \left( \sin^2\frac{\theta}{2}+1\right)
\end{equation}
from Eq.~(\ref{eq:n1a}).
The total probability (\ref{eq:QC-total}) for the one-count process
is then
\begin{equation}
   p^\qc (1) =\frac{3}{2}\gamma^2
\end{equation}
owing to Eq.~(\ref{eq:n1}).
On the contrary, given the one-count process,
the probability density (\ref{eq:QC-conditional}) for
the pre-measurement state $\ket{\psi(a)}$ is
\begin{equation}
 p^\qc(a|1)=\frac{1}{6\pi}\left(
       \sin^2\frac{\theta}{2}+1\right).
\end{equation}
Similarly, given the no-count process,
the probability density for $\ket{\psi(a)}$ is
\begin{equation}
 p^\qc(a|0) \simeq  \frac{1}{4\pi}\left[1-\gamma^2
       \left(\sin^2\frac{\theta}{2}-\frac{1}{2}\right)\right].
\end{equation}
Figure~\ref{fig2} shows these probability densities as
functions of $\theta$ when $\gamma=0.3$.
The one-count process of the quantum counter (\ref{eq:QC-operator})
deforms the probability density to a smoother slope
than that done by the conventional photon counter (\ref{eq:PC-operator}),
not excluding the possibility of $\ket{0}$
owing to the sensitivity to the vacuum state.
Using
\begin{equation}
  \overline{(n_1+1)\log_2 (n_1+1)}
    =2-\frac{3}{4\ln 2},
\end{equation}
we find the information gain (\ref{eq:QC-information})
by the one-count process as
\begin{equation}
    I^\qc(1) =\frac{7}{3}-\frac{1}{2\ln 2}-\log_2 3
       \simeq 0.0270
\label{eq:QC-info1}
\end{equation}
and the mean information gain (\ref{eq:QC-I}) by the measurement as
\begin{equation}
    I^\qc \simeq
  \left(\frac{7}{2}-\frac{3}{4\ln2}-\frac{3}{2}\log_2 3 \right)\gamma^2
   \simeq 0.0405\gamma^2.
\end{equation}
Moreover, the fidelity (\ref{eq:QC-fidelity}) after the one-count process is
\begin{equation}
   F^\qc(1) =\overline{\frac{\sqrt{n_1+1}}{\overline{n_1}+1}
     \bigl|\,c_1^\ast\,c_0\bigr|}
     =\frac{1}{3}\,\mathrm{B}\!\left(\frac{3}{4},\frac{3}{2}\right)
    \simeq 0.320
\label{eq:QC-fide1}
\end{equation}
and the mean fidelity (\ref{eq:QC-F}) after the measurement is
\begin{equation}
 F^\qc \simeq 1-\frac{3}{2}\left[1-\frac{1}{3}\,
       \mathrm{B}\!\left(\frac{3}{4},\frac{3}{2}\right)\right]\gamma^2
    \simeq 1-1.02\gamma^2,
\end{equation}
where $\mathrm{B}(p,q)$ is the beta function.
The reversibilities (\ref{eq:QC-reversibility}) and
(\ref{eq:QC-reversibility0})
of the one-count and no-count processes are then
\begin{align}
  R^\qc(1) &=\frac{2}{3},  \label{eq:QC-reversibility1} \\
  R^\qc(0) &\simeq 1-\frac{1}{2}\gamma^2,
\end{align}
respectively.
Equation~(\ref{eq:QC-reversibility1}) shows that
the one-count process of the quantum counter (\ref{eq:QC-operator})
is physically reversible.
That is, we can, in principle, recover the pre-measurement state
from the post-measurement state with probability $2/3$ on average,
even though it would be difficult
to experimentally implement the reversing measurement
of the quantum counter. 
The reversibility is because of the sensitivity to the vacuum state, namely,
$p^\qc(1|a)>0$ even for $\ket{\psi(a)}=\ket{0}$.
The mean reversibility (\ref{eq:QC-R}) of the measurement is then given by
\begin{equation}
  R^\qc\simeq 1-\gamma^2.
\end{equation}

\section{\label{sec:QPC}QND Photon Counter}
Next, we consider a QND version of the conventional photon counter.
Its measurement operators for one-count
and no-count processes are given by~\cite{UINO92,UeImNa96}
\begin{equation}
 \hat{N}_1=\gamma \hat{a}^\dagger \hat{a}, \qquad
 \hat{N}_0\simeq\hat{I}-\frac{\gamma^2}{2}
     \left(\hat{a}^\dagger \hat{a}\right)^2,
\label{eq:QPC-operator}
\end{equation}
respectively.
This counter neither absorbs nor emits a photon
in both the one-count and no-count processes
through the state reduction (\ref{eq:psim}), thereby
not perturbing the photon-number states $\{\ket{n}\}$
to perform an unsharp QND measurement of photon number.
The measurement operators $\hat{N}_1$ and $\hat{N}_0$
also satisfy the completeness condition (\ref{eq:compcond}), namely,
\begin{equation}
\sum_{m=0, 1} \hcs{\hat{N}_m}\simeq \hat{I},
\end{equation}
up to the order of $\gamma^3$.

A physical model of the QND photon counter is
an indirect measurement described in Sec.~\ref{sec:QM}.
In this case,
the probe is an atom having two degenerate
states $\ket{a}_\text{p}$ and $\ket{b}_\text{p}$
with a transition operator $\hat{\sigma}=\ket{b}_{\text{p}\,\text{p}}\bra{a}$.
The initial state of the atom is the state $\ket{a}_\text{p}$,
and the interaction Hamiltonian between the atom and the photon field is
\begin{equation}
\hat{H}_\text{int}=\hbar g\hat{a}^\dagger \hat{a} \left(\hat{\sigma}
    + \hat{\sigma}^\dagger \right).
\label{eq:QPC-interaction}
\end{equation}
Performing the projective measurement with respect to
the basis $\{\ket{a}_\text{p},\ket{b}_\text{p}\}$,
we recognize that the one-count process has occurred
if the atom is found to be in the state $\ket{b}_\text{p}$
or that the no-count process has occurred
if the atom is found to be still in the state $\ket{a}_\text{p}$.

\subsection{General Model}
To evaluate the amount of information provided
by the QND photon counter (\ref{eq:QPC-operator}),
we consider the set of predefined states $\{\ket{\psi(a)}\}$
described in Sec.~\ref{sec:PC}.
If the pre-measurement state is $\ket{\psi(a)}$,
the probability for the one-count process is
\begin{equation}
 p^\qpc(1|a)  =\bra{\psi(a)} \hat{N}_1^\dagger\hat{N}_1 \ket{\psi(a)}
     = \gamma^2 n_2(a)
\label{eq:QPC-probability}
\end{equation}
owing to Eq.~(\ref{eq:pm}), where
\begin{equation}
   n_2(a) \equiv \sum_n n^2 \left| c_n(a)\right|^2.
\end{equation}
Then, the total probability for the one-count process is given by
\begin{equation}
  p^\qpc(1) =\sum_a  p^\qpc(1|a)\, p(a)=\gamma^2 \overline{n_2}.
\label{eq:QPC-total}
\end{equation}
On the contrary, given the one-count process,
the probability for the pre-measurement state $\ket{\psi(a)}$ is
\begin{equation}
 p^\qpc(a|1) =\frac{p^\qpc(1|a)\, p(a)}{p^\qpc(1)}
   =\frac{n_2(a)}{N\overline{n_2}}.
\label{eq:QPC-conditional}
\end{equation}
Therefore, we obtain the information gain by the one-count process as
\begin{equation}
    I^\qpc(1) =H_0-H^\qpc(1)=\frac{\overline{n_2\log_2 n_2}-
           \overline{n_2}\log_2\overline{n_2}}{\overline{n_2}},
\label{eq:QPC-information}
\end{equation}
where $H^\qpc(1)$ is the Shannon entropy  associated with
the probability distribution (\ref{eq:QPC-conditional}).
This means that our knowledge about the pre-measurement state
increases by $I^\qpc(1)$
when we revise the probability distribution from $p(a)=1/N$
to $p^\qpc(a|1)$ according to the outcome.
On the other hand,
the total probability for the no-count process is
$p^\qpc(0)\simeq 1-\gamma^2 \overline{n_2}$ and
the information gain by the no-count process
is $I^\qpc(0)\simeq 0$ up to the order of $\gamma^3$.
The mean information gain by the measurement thus becomes
\begin{equation}
  I^\qpc=\sum_m p^\qpc(m)\,  I^\qpc(m)\simeq \gamma^2\left(\,
     \overline{n_2\log_2 n_2}-
           \overline{n_2}\log_2\overline{n_2}\,\right).
\label{eq:QPC-I}
\end{equation}

Then, we evaluate the state change owing to the measurement using fidelity.
When the pre-measurement state is $\ket{\psi(a)}$,
the post-measurement state after the one-count process is,
from Eq.~(\ref{eq:psim}),
\begin{equation}
  \ket{\psi(1,a)}^\qpc=\frac{1}{\sqrt{p^\qpc(1|a)}}\, \hat{N}_1 \ket{\psi(a)}
    =\frac{1}{\sqrt{n_2(a)}}\sum_n n\, c_n(a) \, \ket{n},
\end{equation}
with the fidelity to $\ket{\psi(a)}$ being
\begin{equation}
  F^\qpc(1,a) = \bigl|\expv{\psi(a)|\psi(1,a)}^\qpc\bigr|
     =\frac{n_1(a)}{\sqrt{n_2(a)}}.
\end{equation}
Averaging over $a$ with the probability (\ref{eq:QPC-conditional}),
we obtain the fidelity after the one-count process as
\begin{equation}
  F^\qpc(1) =\sum_a p^\qpc(a|1)\,  F^\qpc(1,a)
     = \frac{\overline{\sqrt{n_2}\,n_1}}{\overline{n_2}}.
\label{eq:QPC-fidelity}
\end{equation}
Since the fidelity after the no-count process is $F^\qpc(0) \simeq 1$,
the mean fidelity after the measurement becomes
\begin{equation}
    F^\qpc=\sum_m p^\qpc(m)\,  F^\qpc(m)
   \simeq 1-\gamma^2 \overline{n_2}
     +\gamma^2 \overline{\sqrt{n_2}\,n_1}.
\label{eq:QPC-F}
\end{equation}

Furthermore, we evaluate the reversibility of the measurement using
the maximal successful probability (\ref{eq:maxrecover})
of its reversing measurement.
From the background in Eq.~(\ref{eq:upper}),
$b^\qpc(m)=\inf_{a'} p^\qpc(m|a')$, with
$n_{2\text{i}}=\inf_{a'} n_2(a')$ and $n_{2\text{s}}=\sup_{a'} n_2(a')$,
the reversibilities of the one-count and no-count processes are
\begin{align}
  R^\qpc(1,a) &= \frac{b^\qpc(1)}{p^\qpc(1|a)}
                      = \frac{n_{2\text{i}}}{n_2(a)},   \\
  R^\qpc(0,a) &= \frac{b^\qpc(0)}{p^\qpc(0|a)}
           \simeq \frac{1-\gamma^2n_{2\text{s}}}{1-\gamma^2 n_2(a)},
\end{align}
respectively, if the pre-measurement state is $\ket{\psi(a)}$.
Therefore, they become
\begin{align}
  R^\qpc(1) &= \sum_a p^\qpc(a|1)\, R^\qpc(1,a)
                      = \frac{n_{2\text{i}}}{\,\overline{n_2}\,} ,
  \label{eq:QPC-reversibility}    \\
  R^\qpc(0) &= \sum_a p^\qpc(a|0)\, R^\qpc(0,a)
           \simeq \frac{1-\gamma^2n_{2\text{s}}}{1-\gamma^2 \overline{n_2}},
  \label{eq:QPC-reversibility0}
\end{align}
respectively, if averaged over $a$ with probability (\ref{eq:QPC-conditional}).
The mean reversibility of the measurement is then given by
\begin{equation}
  R^\qpc =\sum_m p^\qpc(m)\,  R^\qpc(m)
   \simeq 1-\gamma^2\left(n_{2\text{s}}-n_{2\text{i}} \right).
\label{eq:QPC-R}
\end{equation}

\subsection{Two-state Model}
\begin{figure}
\begin{center}
\includegraphics[scale=0.7]{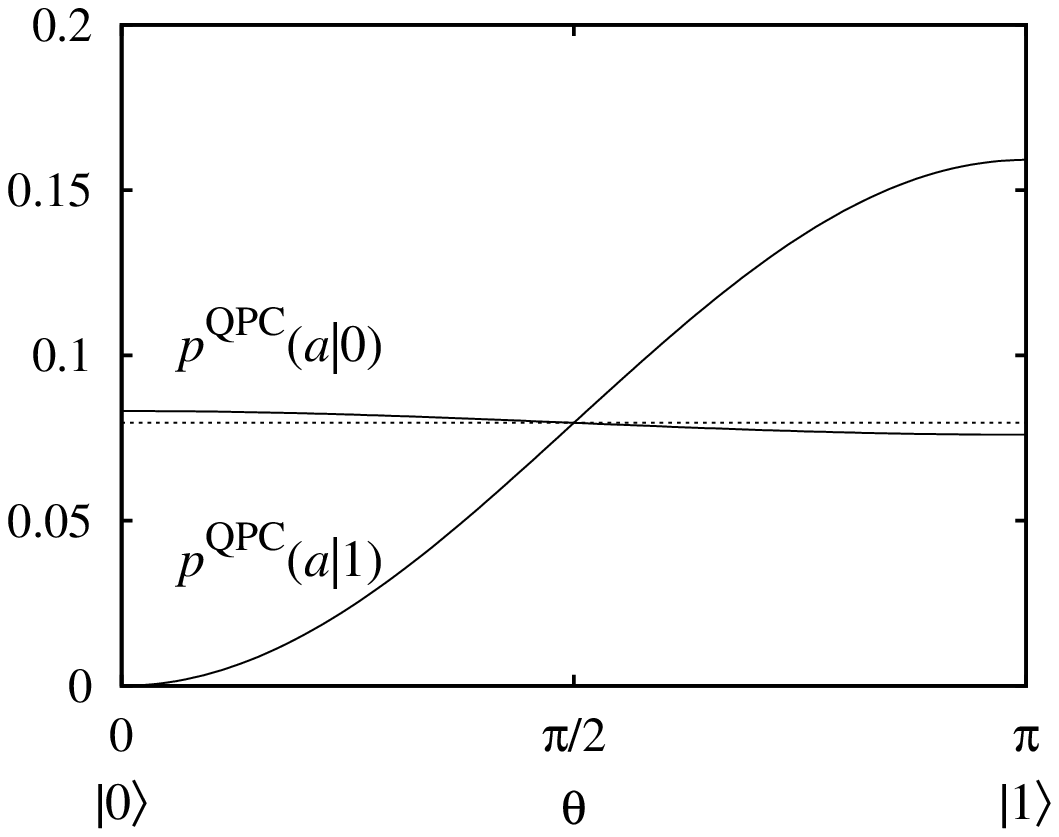}
\end{center}
\caption{\label{fig3}
Probability density for the pre-measurement state $\ket{\psi(a)}$
conditioned by the one-count process $p^\qpc(a|1)$
and that conditioned by the no-count process $p^\qpc(a|0)$
as functions of $\theta$ when $\gamma=0.3$.
The dotted line indicates the initial probability density $p(a)=1/4\pi$.}
\end{figure}

We again consider the situation discussed
in Sec.~\ref{sec:PC} as an example.
If the pre-measurement state is $\ket{\psi(a)}$,
the probability (\ref{eq:QPC-probability}) 
for the one-count process is
\begin{equation}
   p^\qpc(1|a) = \gamma^2 \sin^2\frac{\theta}{2},
\end{equation}
since using Eq.~(\ref{eq:psi}) we have
\begin{equation}
   n_2(a)=\sum_{n=0,1} n^2 \left| c_n(a)\right|^2
     =\sin^2\frac{\theta}{2}.
\end{equation}
Note that $n_2(a)=n_1(a)$
in this two-state model since $n^2=n$ for $n=0,1$.
Therefore, from Eq.~(\ref{eq:n1}), we obtain
\begin{equation}
\overline{n_2}=\frac{1}{2}.
\end{equation}
The total probability (\ref{eq:QPC-total}) for the one-count process
is thus given by
\begin{equation}
   p^\qpc (1) =\frac{1}{2}\gamma^2.
\end{equation}
On the contrary, given the one-count process,
the probability density (\ref{eq:QPC-conditional})
for $\ket{\psi(a)}$ is
\begin{equation}
 p^\qpc(a|1)=\frac{1}{2\pi}
       \sin^2\frac{\theta}{2},
\end{equation}
and the corresponding probability density for the no-count process is
\begin{equation}
 p^\qpc(a|0) \simeq  \frac{1}{4\pi}\left[1-\gamma^2
       \left(\sin^2\frac{\theta}{2}-\frac{1}{2}\right)\right].
\end{equation}
These probability densities are shown in Fig.~\ref{fig3},
which is the same form as Fig.~\ref{fig1}
owing to $n_2(a)=n_1(a)$ in this two-state model.
Since we have
\begin{equation}
\overline{n_2\log_2 n_2}=-\frac{1}{4\ln2}
\end{equation}
as in Eq.~(\ref{eq:n1logn1}),
the information gain (\ref{eq:QPC-information})
by the one-count process becomes
\begin{equation}
  I^\qpc(1) =1-\frac{1}{2\ln 2}
       \simeq 0.279
\label{eq:QPC-info1}
\end{equation}
and the mean information gain (\ref{eq:QPC-I}) by the measurement becomes
\begin{equation}
  I^\qpc\simeq \left(\frac{1}{2}-\frac{1}{4\ln2} \right)\gamma^2
   \simeq 0.139\gamma^2.
\end{equation}
On the other hand,
the fidelity (\ref{eq:QPC-fidelity}) after the one-count process is
\begin{equation}
   F^\qpc(1)= \frac{\overline{\sqrt{n_2}\,n_1}}{\overline{n_2}}
       =\frac{4}{5}
\label{eq:QPC-fide1}
\end{equation}
and the mean fidelity (\ref{eq:QPC-F}) after the measurement is
\begin{equation}
  F^\qpc \simeq 1-\frac{1}{10}\gamma^2.
\end{equation}
In addition,
since we have $n_{2\text{i}}=\inf_{a'} n_2(a')=0$
with $\ket{\psi(a')}=\ket{0}$
and $n_{2\text{s}}=\sup_{a'} n_2(a')=1$
with $\ket{\psi(a')}=\ket{1}$,
the reversibilities (\ref{eq:QPC-reversibility}) and
(\ref{eq:QPC-reversibility0})
of the one-count and no-count processes are given by
\begin{align}
  R^\qpc(1) &=0,  \label{eq:QPC-reversibility1} \\
  R^\qpc(0) &\simeq 1-\frac{1}{2}\gamma^2,
\end{align}
respectively.
As shown in Eq.~(\ref{eq:QPC-reversibility1}),
the one-count process of the QND photon counter (\ref{eq:QPC-operator})
is not physically reversible.
We cannot recover the pre-measurement state
from the post-measurement state
as in the case of the conventional photon counter (\ref{eq:PC-operator}).
Note that the QND photon counter is not sensitive to the vacuum state
owing to $p^\qpc(1|a) =0$ for $\ket{\psi(a)}=\ket{0}$.
The mean reversibility (\ref{eq:QPC-R}) of the measurement is thus
\begin{equation}
  R^\qpc\simeq 1-\gamma^2.
\end{equation}

\section{\label{sec:QQC}QND Quantum Counter}
In this section, we propose a novel type of photon counter,
that is, a QND version of the quantum counter,
whose measurement operators for one-count
and no-count processes are written as
\begin{equation}
 \hat{Q}_1=\gamma \hat{a}\hat{a}^\dagger , \qquad
 \hat{Q}_0\simeq\hat{I}-\frac{\gamma^2}{2}
     \left( \hat{a}\hat{a}^\dagger\right)^2,
\label{eq:QQC-operator}
\end{equation}
respectively.
This counter performs a reversible QND measurement of photon number
because it is sensitive not only to photons
but also to the vacuum state
without perturbing the photon-number states $\{\ket{n}\}$.
Of course,
the measurement operators $\hat{Q}_1$ and $\hat{Q}_0$
satisfy the completeness condition (\ref{eq:compcond}),
\begin{equation}
\sum_{m=0, 1} \hcs{\hat{Q}_m}\simeq \hat{I},
\end{equation}
up to the order of $\gamma^3$.
A physical model of the QND quantum counter is
similar to that of the QND photon counter described
in Sec.~\ref{sec:QPC}.
The only difference is that
the interaction Hamiltonian between the atom and the photon field
is now
\begin{equation}
\hat{H}_\text{int}=\hbar g \hat{a}\hat{a}^\dagger \left(\hat{\sigma}
    + \hat{\sigma}^\dagger \right),
\end{equation}
instead of Eq.~(\ref{eq:QPC-interaction}).

\subsection{General Model}
The amount of information provided
by the QND quantum counter (\ref{eq:QQC-operator})
is evaluated using the set of predefined states $\{\ket{\psi(a)}\}$
as in Sec.~\ref{sec:PC}.
If the pre-measurement state is $\ket{\psi(a)}$,
the probability for the one-count process is
\begin{equation}
 p^\qqc(1|a) =\bra{\psi(a)} \hat{Q}_1^\dagger\hat{Q}_1 \ket{\psi(a)}
     = \gamma^2 n_3(a),
\label{eq:QQC-probability}
\end{equation}
according to Eq.~(\ref{eq:pm}), where
\begin{equation}
   n_3(a) \equiv \sum_n (n+1)^2 \left| c_n(a)\right|^2.
\end{equation}
The total probability for the one-count process is thus
\begin{equation}
  p^\qqc(1) =\sum_a  p^\qqc(1|a)\, p(a)
     =\gamma^2 \overline{n_3}.
\label{eq:QQC-total}
\end{equation}
On the contrary, given the one-count process,
the probability for the pre-measurement state $\ket{\psi(a)}$ is
\begin{equation}
 p^\qqc(a|1) =\frac{p^\qqc(1|a)\, p(a)}{p^\qqc(1)}
   =\frac{n_3(a)}{N\overline{n_3}}.
\label{eq:QQC-conditional}
\end{equation}
Calculating the Shannon entropy $H^\qqc(1)$ associated with
the probability distribution (\ref{eq:QQC-conditional}),
we find the information gain by the one-count process as
\begin{equation}
   I^\qqc(1) =H_0-H^\qqc(1)= \frac{\overline{n_3\log_2 n_3}-
       \overline{n_3}\log_2 \overline{n_3}}{\overline{n_3}},
\label{eq:QQC-information}
\end{equation}
which quantifies 
the increase in
our knowledge about the pre-measurement state
when the probability distribution $p(a)=1/N$
is revised to $p^\qqc(a|1)$ according to the outcome.
In a similar way, we obtain
the total probability for the no-count process as
$p^\qqc(0)\simeq 1-\gamma^2 \overline{n_3}$
and the information gain by the no-count process
as $I^\qqc(0)\simeq 0$ up to the order of $\gamma^3$.
The mean information gain by the measurement is thus
\begin{equation}
 I^\qqc =\sum_m p^\qqc(m)\,  I^\qqc(m)
    \simeq \gamma^2\left(\,\overline{n_3\log_2 n_3}-
  \overline{n_3}\log_2 \overline{n_3} \,\right).
\label{eq:QQC-I}
\end{equation}

Furthermore,
the state change owing to the measurement is evaluated by fidelity.
According to Eq.~(\ref{eq:psim}),
when the pre-measurement state is $\ket{\psi(a)}$,
the post-measurement state after the one-count process is
\begin{equation}
  \ket{\psi(1,a)}^\qqc 
    = \frac{1}{\sqrt{p^\qqc(1|a)}}\, \hat{Q}_1 \ket{\psi(a)} 
      =\frac{1}{\sqrt{n_3(a)}}\sum_n (n+1)\, c_n(a) \, \ket{n}.
\end{equation}
Therefore, the fidelity to $\ket{\psi(a)}$ becomes
\begin{equation}
  F^\qqc(1,a) = \bigl|\expv{\psi(a)|\psi(1,a)}^\qqc\bigr|
     =\frac{n_1(a)+1}{\sqrt{n_3(a)}}
\end{equation}
after the one-count process.
Averaging over $a$ with the probability (\ref{eq:QQC-conditional}),
we obtain the fidelity after the one-count process as
\begin{equation}
  F^\qqc(1) =\sum_a p^\qqc(a|1)\,  F^\qqc(1,a)
     = \frac{\overline{\sqrt{n_3}\,\left(n_1+1\right)}}{\overline{n_3}}.
\label{eq:QQC-fidelity}
\end{equation}
Since the fidelity after the no-count process is
$F^\qqc(0) \simeq 1$,
the mean fidelity after the measurement is given by
\begin{equation}
    F^\qqc =\sum_m p^\qqc(m)\,  F^\qqc(m) 
    \simeq 1-\gamma^2 \overline{n_3}
     +\gamma^2 \overline{\sqrt{n_3}\,\left(n_1+1\right)}.
\label{eq:QQC-F}
\end{equation}

Finally, the reversibility of the measurement is evaluated by
the maximal successful probability (\ref{eq:maxrecover})
of its reversing measurement.
If the pre-measurement state is $\ket{\psi(a)}$,
the reversibilities of the one-count and no-count processes
are respectively given by
\begin{align}
  R^\qqc(1,a) &= \frac{b^\qqc(1)}{p^\qqc(1|a)}
         = \frac{n_{3\text{i}}}{n_3(a)}, \\
  R^\qqc(0,a) &= \frac{b^\qqc(0)}{p^\qqc(0|a)}
     \simeq\frac{1-\gamma^2 n_{3\text{s}}}{1-\gamma^2 n_3(a)},
\end{align}
where we have used the background $b^\qqc(m)=\inf_{a'} p^\qqc(m|a')$
defined in Eq.~(\ref{eq:upper}), and
$n_{3\text{i}}=\inf_{a'} n_3(a')$ and $n_{3\text{s}}=\sup_{a'} n_3(a')$.
Averaged over $a$ with the probability (\ref{eq:QQC-conditional}),
they become
\begin{align}
  R^\qqc(1) &= \sum_a p^\qqc(a|1)\, R^\qqc(1,a)
     = \frac{n_{3\text{i}}}{\,\overline{n_3}\,},
    \label{eq:QQC-reversibility} \\
  R^\qqc(0) &= \sum_a p^\qqc(a|0)\, R^\qqc(0,a) 
  \simeq\frac{1-\gamma^2 n_{3\text{s}}}{1-\gamma^2 \overline{n_3} }.
   \label{eq:QQC-reversibility0}
\end{align}
Thus, the mean reversibility of the measurement is
\begin{equation}
R^\qqc =\sum_m p^\qqc(m)\,  R^\qqc(m)
   \simeq 1-\gamma^2\left( n_{3\text{s}}-n_{3\text{i}}\right).
\label{eq:QQC-R}
\end{equation}

\subsection{Two-state Model}
\begin{figure}
\begin{center}
\includegraphics[scale=0.7]{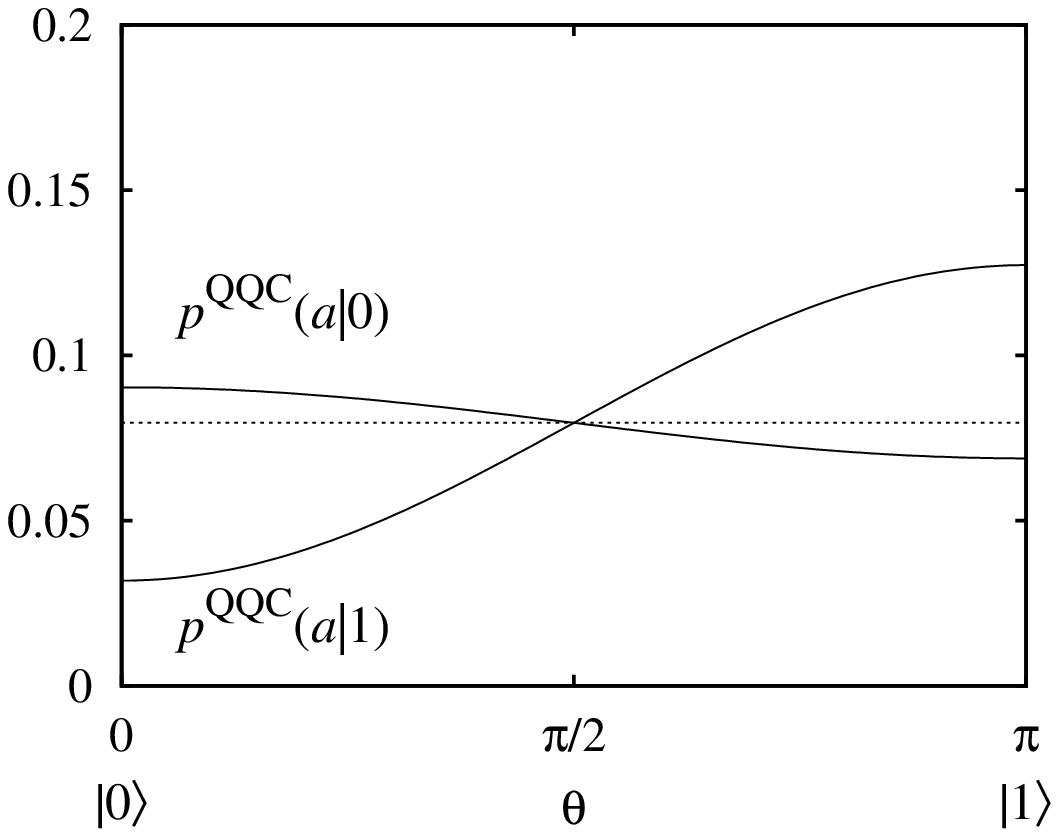}
\end{center}
\caption{\label{fig4}
Probability density for the pre-measurement state $\ket{\psi(a)}$
conditioned by the one-count process $p^\qqc(a|1)$
and that conditioned by the no-count process $p^\qqc(a|0)$
as functions of $\theta$ when $\gamma=0.3$.
The dotted line indicates the initial probability density $p(a)=1/4\pi$.}
\end{figure}

We consider the situation discussed
in Sec.~\ref{sec:PC} as an example.
If the pre-measurement state is $\ket{\psi(a)}$,
the probability (\ref{eq:QQC-probability}) 
for the one-count process is
\begin{equation}
   p^\qqc(1|a) = \gamma^2 \left(3 \sin^2\frac{\theta}{2}+1\right),
\end{equation}
since
\begin{equation}
   n_3(a)=\sum_{n=0,1} (n+1)^2 \left| c_n(a)\right|^2
     =3\sin^2\frac{\theta}{2}+1
\end{equation}
from Eq.~(\ref{eq:psi}).
Using
\begin{equation}
  \overline{n_3}=\frac{5}{2},
\end{equation}
we find
the total probability (\ref{eq:QQC-total}) for the one-count process as
\begin{equation}
   p^\qqc (1) =\frac{5}{2}\gamma^2.
\end{equation}
On the contrary, given the one-count process,
the probability density (\ref{eq:QQC-conditional}) for
the pre-measurement state $\ket{\psi(a)}$ is
\begin{equation}
  p^\qqc(a|1) =\frac{1}{10\pi}\left(3\sin^2\frac{\theta}{2}+1\right),
\end{equation}
while the corresponding probability density for the no-count process is
\begin{equation}
 p^\qqc(a|0) \simeq \frac{1}{4\pi}\left[1-3\gamma^2
       \left(\sin^2\frac{\theta}{2}-\frac{1}{2}\right)\right].
\end{equation}
These probability densities are shown in Fig.~\ref{fig4}.
Note that the possibility of $\ket{0}$ is not
excluded by the one-count process, but
it is less than that in the case of the quantum counter.
Since
\begin{equation}
 \overline{n_3\log_2 n_3} =\frac{16}{3}-\frac{5}{4\ln 2},
\end{equation}
the information gain (\ref{eq:QQC-information})
by the one-count process is
\begin{equation}
   I^\qqc(1) =\frac{47}{15}-\frac{1}{2\ln 2}-\log_2 5
       \simeq 0.0901
\label{eq:QQC-info1}
\end{equation}
and the mean information gain (\ref{eq:QQC-I}) by the measurement is
\begin{equation}
  I^\qqc\simeq
  \left(\frac{47}{6}-\frac{5}{4\ln2}-\frac{5}{2}\log_2 5 \right)\gamma^2
   \simeq 0.225\gamma^2.
\end{equation}
Moreover,
the fidelity (\ref{eq:QQC-fidelity}) after the one-count process becomes
\begin{equation}
    F^\qqc(1)= 
    \frac{\overline{\sqrt{n_3}\,\left(n_1+1\right)}}{\overline{n_3}}
       =\frac{652}{675}
\label{eq:QQC-fide1}
\end{equation}
and the mean fidelity (\ref{eq:QQC-F}) after the measurement becomes
\begin{equation}
   F^\qqc \simeq 1-\frac{23}{270}\gamma^2.
\end{equation}
On the other hand,
since we have $n_{3\text{i}}=\inf_{a'} n_3(a')=1$ with $\ket{\psi(a')}=\ket{0}$
and $n_{3\text{s}}=\sup_{a'} n_3(a')=4$ with $\ket{\psi(a')}=\ket{1}$,
the reversibilities (\ref{eq:QQC-reversibility})
and (\ref{eq:QQC-reversibility0}) of
the one-count and no-count processes are given by
\begin{align}
  R^\qqc(1) &=\frac{2}{5},  \label{eq:QQC-reversibility1} \\
  R^\qqc(0) &\simeq 1-\frac{3}{2}\gamma^2,
\end{align}
respectively.
As in Eq.~(\ref{eq:QQC-reversibility1}),
the one-count process of the QND quantum counter (\ref{eq:QQC-operator})
is physically reversible 
because of the sensitivity to the vacuum state, namely,
$p^\qqc(1|a)>0$ even for $\ket{\psi(a)}=\ket{0}$.
Thus, we can recover the pre-measurement state
from the post-measurement state
as in the case of the quantum counter (\ref{eq:QC-operator}).
However, in the QND quantum counter,
the successful recovery occurs with probability $2/5$
on average, which is less than
that in the quantum counter, Eq.~(\ref{eq:QC-reversibility1}).
The mean reversibility (\ref{eq:QQC-R}) of the measurement is then
\begin{equation}
  R^\qqc\simeq 1-3\gamma^2.
\end{equation}

\section{\label{sec:summary}Summary and Discussion}
\begin{figure}
\begin{center}
\includegraphics[scale=0.6]{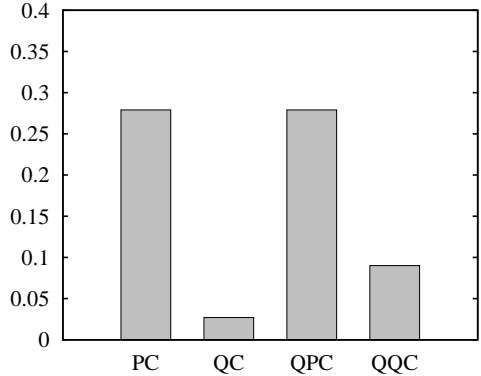}
\end{center}
\caption{\label{fig5}
Information gain by the one-count process
in the two-state model.
PC, QC, QPC, and QQC denote
the conventional photon counter, quantum counter, QND photon counter,
and QND quantum counter, respectively.}
\end{figure}

\begin{figure}
\begin{center}
\includegraphics[scale=0.6]{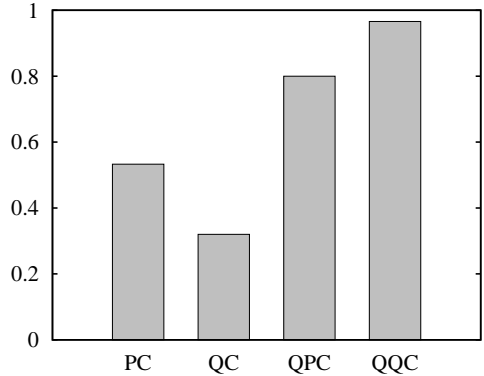}
\end{center}
\caption{\label{fig6}
Fidelity after the one-count process in the two-state model.
PC, QC, QPC, and QQC denote
the conventional photon counter, quantum counter, QND photon counter,
and QND quantum counter, respectively.}
\end{figure}

\begin{figure}
\begin{center}
\includegraphics[scale=0.6]{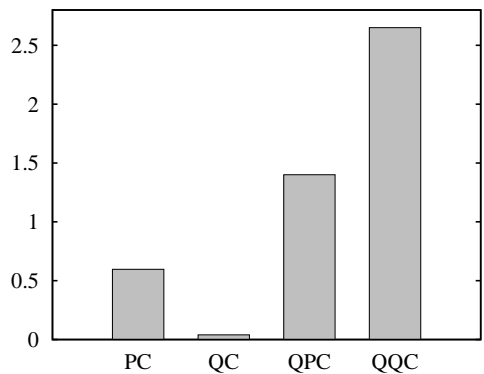}
\end{center}
\caption{\label{fig7}
Efficiency of the one-count process in the two-state model.
PC, QC, QPC, and QQC denote
the conventional photon counter, quantum counter, QND photon counter,
and QND quantum counter, respectively.}
\end{figure}

\begin{figure}
\begin{center}
\includegraphics[scale=0.6]{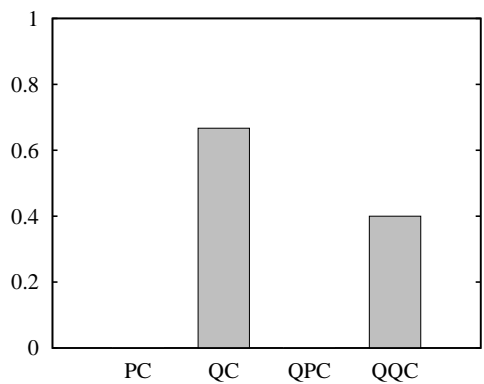}
\end{center}
\caption{\label{fig8}
Physical reversibility of the one-count process
in the two-state model.
PC, QC, QPC, and QQC denote
the conventional photon counter, quantum counter, QND photon counter,
and QND quantum counter, respectively.}
\end{figure}

We investigated four types of photon counters:
conventional photon counter, quantum counter, QND photon counter,
and QND quantum counter.
For each counter, we calculated
information gain, fidelity, and physical reversibility,
assuming that a photon field to be measured is
in an arbitrary superposition of the vacuum state $\ket{0}$
and the one-photon state $\ket{1}$.
Figure~\ref{fig5} displays
the information gain by the one-count process of each counter,
namely, Eqs.~(\ref{eq:PC-info1}), (\ref{eq:QC-info1}),
(\ref{eq:QPC-info1}), and (\ref{eq:QQC-info1}).
The conventional photon counter and the QND photon counter
provide the same amount of information
in the two-state model.
However, if the photon field is in
an arbitrary superposition of the three states
$\ket{0}$, $\ket{1}$, and $\ket{2}$,
a numerical calculation shows that
the QND photon counter provides more information than
the conventional photon counter.
Therefore, the QND photon counter
has an advantage in terms of information gain.
In contrast, the quantum counter provides
about $10$ times less information
than the QND photon counter.
On the other hand,
Fig.~\ref{fig6} displays
the fidelity after the one-count process for each counter,
namely, Eqs.~(\ref{eq:PC-fide1}), (\ref{eq:QC-fide1}),
(\ref{eq:QPC-fide1}), and (\ref{eq:QQC-fide1}).
The QND versions change
the state of the photon field less than
that changed by their original versions.
In particular,
the QND quantum counter 
almost retains the state of photon field,
compared with the quantum counter.
To emphasize this property,
we define an efficiency of counter by
the ratio of information gain to fidelity loss, e.g.,
for the conventional photon counter
\begin{equation}
  E^\pc(1)\equiv \frac{I^\pc(1)}{1-F^\pc(1)},
\end{equation}
and so on.
Then, the QND quantum counter has approximately
twice the efficiency of
the QND photon counter, as shown in Fig.~\ref{fig7}.
Figure~\ref{fig8} displays
the physical reversibility of the one-count process of each counter,
namely, Eqs.~(\ref{eq:PC-reversibility1}), (\ref{eq:QC-reversibility1}),
(\ref{eq:QPC-reversibility1}), and (\ref{eq:QQC-reversibility1}).
We can see that the quantum counter is the most reversible counter,
while the conventional photon counter and the QND photon counter
are irreversible.

Our results suggest that the reversibility of a counter
tends to decrease the amount of information obtained by the counter.
A similar result was shown~\cite{Ban01}
using reversible spin-$1/2$ measurement~\cite{Royer94}.
However, the reversibility of a counter does not necessarily decrease
the state change caused by the counter.
In fact, the quantum counter has the highest reversibility
and provides the smallest amount of information
but changes the state of the photon field most.
This is because of
a unitary part of the measurement operator~\cite{FucJac01,TerUed07b}.
Note that the measurement operator $\hat{L}_1$ in Eq.~(\ref{eq:QC-operator})
could be written by polar decomposition as
\begin{equation}
  \hat{L}_1=\gamma\, \hat{U} \sqrt{\hat{a}\hat{a}^\dagger},
\end{equation}
where $\hat{U}$ is a unitary operator and
$\sqrt{\hat{a}\hat{a}^\dagger}$ is a non-negative operator,
as long as the Hilbert space of the photon field is truncated
to finite dimensions~\cite{Fujika95}, as in the two-state model.
The unitary part $\hat{U}$ causes an additional state change
after the raw measurement $\sqrt{\hat{a}\hat{a}^\dagger}$,
leaving the information gain and physical reversibility invariant.
Therefore, the highest reversibility with the least information
does not imply high fidelity in the quantum counter.
Among the other counters,
the conventional photon counter (\ref{eq:PC-operator})
also has such a unitary part, while
the remaining two counters do not have a unitary part.
A general theory on the relations among information, fidelity, and
reversibility would be developed elsewhere.

We could implement the QND quantum counter proposed
in Sec.~\ref{sec:QQC} using a joint measurement.
Consider performing the first measurement by the quantum counter
and the second measurement by the conventional photon counter.
If both the counters detect photons,
the total process of the joint measurement is equivalent to
the one-count process of the QND quantum counter
because of
\begin{equation}
  \hat{M}_1\hat{L}_1 \propto \hat{Q}_1
\end{equation}
from Eqs.~(\ref{eq:PC-operator}), (\ref{eq:QC-operator}),
and (\ref{eq:QQC-operator}).
The joint measurement is thus an implementation of
the QND quantum counter,
even though there are four possible outcomes.
Note that this implementation is an example of the
Hermitian conjugate measurement scheme~\cite{TerUed07b},
since the second measurement by the conventional photon counter is a
Hermitian conjugate measurement
of the first measurement by the quantum counter
owing to $\hat{M}_1\propto\hat{L}_1^\dagger$.
Therefore, the second measurement
cancels the unitary part $\hat{U}$
of the measurement operator $\hat{L}_1$,
thereby increasing the fidelity and information gain
to the extent of a single measurement by the QND quantum counter.

\section*{Acknowledgments}
The author thanks M. Ueda for helpful comments.
This research was supported by a Grant-in-Aid
for Scientific Research (Grant No.~20740230) from
the Ministry of Education, Culture, Sports,
Science and Technology of Japan.


\begin{thebibliography}{10}

\bibitem{NieChu00}
M.~A. Nielsen and I.~L. Chuang, {\em Quantum Computation and Quantum
  Information} (Cambridge University Press, Cambridge, 2000).

\bibitem{BenBra84}
C.~H. Bennett and G. Brassard,  in {\em Proceedings of IEEE International
  Conference on Computers, Systems and Signal Processing, Bangalore, India}
  (IEEE, New York, 1984), pp.\ 175--179.

\bibitem{Ekert91}
A.~K. Ekert, Phys. Rev. Lett. {\bf 67},  661  (1991).

\bibitem{Bennet92}
C.~H. Bennett, Phys. Rev. Lett. {\bf 68},  3121  (1992).

\bibitem{BeBrMe92}
C.~H. Bennett, G. Brassard, and N.~D. Mermin, Phys. Rev. Lett. {\bf 68},  557
  (1992).

\bibitem{UedKit92}
M. Ueda and M. Kitagawa, Phys. Rev. Lett. {\bf 68},  3424  (1992).

\bibitem{UeImNa96}
M. Ueda, N. Imoto, and H. Nagaoka, Phys. Rev. A {\bf 53},  3808  (1996).

\bibitem{LanLif77}
L.~D. Landau and E.~M. Lifshitz, {\em Quantum Mechanics (Non-Relativistic
  Theory)}, 3rd  ed. (Butterworth-Heinemann, Oxford, 1977).

\bibitem{Ueda97}
M. Ueda, in {\em Frontiers in Quantum Physics: Proceedings of the
  International Conference on Frontiers in Quantum Physics,
  Kuala Lumpur, Malaysia, 1997},
  edited by S.~C. Lim, R. Abd-Shukor, and K.~H. Kwek
  (Springer-Verlag, Singapore, 1999), pp.\ 136--144.

\bibitem{Imamog93}
A. Imamo{\=g}lu, Phys. Rev. A {\bf 47},  R4577  (1993).

\bibitem{Royer94}
A. Royer, Phys. Rev. Lett. {\bf 73},  913  (1994);
{\bf 74}, 1040(E) (1995).

\bibitem{TerUed05}
H. Terashima and M. Ueda, Phys. Rev. A {\bf 74},  012102  (2006).

\bibitem{KorJor06}
A.~N. Korotkov and A.~N. Jordan, Phys. Rev. Lett. {\bf 97},  166805  (2006).

\bibitem{TerUed07}
H. Terashima and M. Ueda, Phys. Rev. A {\bf 75},  052323  (2007).

\bibitem{SuAlZu09}
Q. Sun, M. Al-Amri, and M.~S. Zubairy, Phys. Rev. A {\bf 80},  033838  (2009).

\bibitem{XuZho10}
Y.-Y. Xu and F. Zhou, Commun. Theor. Phys. {\bf 53},  469  (2010).

\bibitem{KoaUed99}
M. Koashi and M. Ueda, Phys. Rev. Lett. {\bf 82},  2598  (1999).

\bibitem{TerUed03}
H. Terashima and M. Ueda, Int. J. Quantum Inf. {\bf 3},  633
  (2005).

\bibitem{KNABHL08}
N. Katz, M. Neeley, M. Ansmann, R.~C. Bialczak, M. Hofheinz, E. Lucero,
A. O'Connell, H. Wang, A.~N. Cleland, J.~M. Martinis, and A.~N. Korotkov,
Phys. Rev. Lett. {\bf 101},  200401  (2008).

\bibitem{KCRK09}
Y.-S. Kim, Y.-W. Cho, Y.-S. Ra, and Y.-H. Kim, Opt. Express {\bf 17},  11978
  (2009).

\bibitem{MabZol96}
H. Mabuchi and P. Zoller, Phys. Rev. Lett. {\bf 76},  3108  (1996).

\bibitem{NieCav97}
M.~A. Nielsen and C.~M. Caves, Phys. Rev. A {\bf 55},  2547  (1997).

\bibitem{TerUed07b}
H. Terashima and M. Ueda, Phys. Rev. A {\bf 81},  012110  (2010).

\bibitem{Ban01}
M. Ban, J. Phys. A: Math. Gen. {\bf 34},  9669  (2001).

\bibitem{DArian03}
G.~M. D'Ariano, Fortschr. Phys. {\bf 51},  318  (2003).

\bibitem{Bloemb59}
N. Bloembergen, Phys. Rev. Lett. {\bf 2},  84  (1959).

\bibitem{Mandel66}
L. Mandel, Phys. Rev. {\bf 152},  438  (1966).

\bibitem{BraKha96}
V.~B. Braginsky and F.~Y. Khalili, Rev. Mod. Phys. {\bf 68},  1  (1996),
and references therein.

\bibitem{Uhlman76}
A. Uhlmann, Rep. Math. Phys. {\bf 9},  273  (1976).

\bibitem{DavLew70}
E.~B. Davies and J.~T. Lewis, Commun. Math. Phys. {\bf 17},  239  (1970).

\bibitem{SriDav81}
M.~D. Srinivas and E.~B. Davies, Opt. Acta {\bf 28},  981  (1981).

\bibitem{UeImOg90}
M. Ueda, N. Imoto, and T. Ogawa, Phys. Rev. A {\bf 41},  3891  (1990).

\bibitem{UNSTN04}
K. Usami, Y. Nambu, B.-S. Shi, A. Tomita, and K. Nakamura,
Phys. Rev. Lett. {\bf 92},  113601  (2004);
K. Usami, A. Tomita, and K. Nakamura, Int. J. Quantum Inf. {\bf 2},  101
  (2004).

\bibitem{UINO92}
M. Ueda, N. Imoto, H. Nagaoka, and T. Ogawa, Phys. Rev. A {\bf 46},  2859
  (1992).

\bibitem{FucJac01}
C.~A. Fuchs and K. Jacobs, Phys. Rev. A {\bf 63},  062305  (2001).

\bibitem{Fujika95}
In the infinite-dimensional Hilbert space spanned by
all the photon-number states $\{\ket{n}\}$ with $n=0,1,2,\ldots$,
the creation operator $\hat{a}^\dagger$
and annihilation operator $\hat{a}$
do not have the polar decomposition:
K. Fujikawa, Phys. Rev. A {\bf 52},  3299  (1995).

\end{thebibliography}

\end{document}